\documentclass[a4paper,11pt]{article}

\usepackage{jinstpub}
\usepackage[utf8]{inputenc}

\usepackage{amsmath}
\usepackage{amssymb}
\usepackage{xspace}
\usepackage{upgreek}
\usepackage{subcaption}
\usepackage{graphicx}
\usepackage{makecell}
\usepackage[page,toc,titletoc,title]{appendix}
\usepackage[subfigure]{tocloft}
\usepackage[section]{placeins} 
\usepackage[export]{adjustbox}
\usepackage{comment}
\usepackage{booktabs}
\usepackage{hyperref}
\usepackage{enumitem}

\usepackage{siunitx}
\DeclareSIUnit\sq{\ensuremath{\Box}} 

\usepackage[rawfloats=true]{floatrow}

\usepackage{xcolor}

\usepackage{lineno}

\title{Pixel response characterization of the ARCADIA Fully Depleted MAPS}

\author[a,f]    {C. Pantouvakis}
\author[a,f]    {M. Rignanese}
\author[k]      {T. Zenger}
\author[b,f]    {S. Ciarlantini}
\author[a,f]    {A. Zingaretti}
\author[f]      {P. Azzi}
\author[b,f]    {C. Bonini}
\author[a,f,m]  {D. Chiappara}
\author[a,f]    {S. Mattiazzo}
\author[a,f]    {D. Pantano}
\author[c,f]    {J. Wyss}
\author[k]      {A. Apresyan}
\author[f,k]    {N. Bacchetta}
\author[k]      {L. Bolla}
\author[n]      {A. Hayrapetyan}
\author[k]      {C. Pena}
\author[k]      {N. Salvador}
\author[k,l]    {S. Xie}
\author[k]      {I. Zoi}
\author[h]      {D. Falchieri}
\author[g]      {S. Garbolino}
\author[d,i]    {L. Pancheri}
\author[g]      {A. Paternò}
\author[g]      {A. Rivetti}
\author[g]      {M. Rolo}
\author[e,j]    {R. Santoro}
\author[a,f,1]  {P. Giubilato\note{Corresponding author}}

\affiliation[a]{University of Padova, Department of Physics and Astronomy "G. Galilei", Padova, Italy}
\affiliation[b]{University of Padova, Centro di ateneo di Studi e Attività Spaziali CISAS "G. Colombo", Padova, Italy}
\affiliation[c]{University of Cassino and Southern Lazio, DICEM, Cassino, Italy}
\affiliation[d]{University of Trento, Department of Industrial Engineering, Trento, Italy}
\affiliation[e]{University of Insubria, Department of Science and Technology, Varese, Italy}
\affiliation[f]{Istituto Nazionale di Fisica Nucleare, sezione di Padova, Padova, Italy}
\affiliation[g]{Istituto Nazionale di Fisica Nucleare, sezione di Torino, Torino, Italy}
\affiliation[h]{Istituto Nazionale di Fisica Nucleare, sezione di Bologna, Bologna, Italy}
\affiliation[i]{Istituto Nazionale di Fisica Nucleare TIFPA, Trento, Italy}
\affiliation[j]{Istituto Nazionale di Fisica Nucleare, sezione di Milano, Milano, Italy}
\affiliation[k]{Fermi National Accelerator Laboratory, Batavia, IL, USA}
\affiliation[l]{California Institute of Technology, Pasadena, CA, USA}
\affiliation[m]{Conseil Européen pour la Recherche Nucléaire (CERN), Genève, Switzerland}
\affiliation[n]{A. I. Alikhanyan National Science Laboratory (Yerevan Physics Institute), Yerevan, Armenia}

\emailAdd{piero.giubilato@unipd.it}

\abstract{Monolithic Active Pixel Sensors (MAPS) achieved widespread use in several scientific applications, thanks to their properties, such as low material budget and high granularity. The ARCADIA INFN project developed a Fully-Depleted MAPS (FD-MAPS), using a modified LFoundry 110 nm CIS process. This work presents the first laboratory characterization of the ARCADIA MD3 prototype. Measurements include threshold uniformity studies using both test-pulse injection and a $^{55}$Fe source, as well as threshold and noise calibration achieved thanks to monochromatic X-ray sources. Ultimately, charge-collection efficiency is evaluated using an infrared laser setup.}
    
\keywords{Solid state detectors; Particle tracking detectors; Performance of High Energy Physics Detectors; X-ray detectors}

\begin{document}
\maketitle
\flushbottom


\section{Introduction}
\label{sec:intro}

Monolithic Active Pixel Sensors (MAPS) are solid-state sensors where the sensing volume and the front-end electronics are integrated within the same silicon by means of standard microelectronic manufacturing processes. This tight integration minimizes parasitic interconnection between the sensor and the front-end electronics, leading to excellent power figures, high granularity, and low material budget. The absence of any post-production processing (e.g. bump bonding) and the availability of industrial lithography processes for high volume production runs, make MAPS an ideal choice whenever low-power, high-resolution, and/or large instrumented area are foreseen. \\
The first and second generations of MAPS, exemplified by the MIMOSA chip \cite{MIMOSA} and the ALPIDE chip developed for the ALICE ITS2 upgrade \cite{ALPIDE}, collect the charge generated in their sensing volume primarily by diffusion. Diffusion is a "slow" process, taking tens of nanoseconds to complete within the typical volumes of MAPS pixel cells, leading to relatively poor timing performance \cite{diffusion, timing-performance}. Furthermore, it hampers the device radiation tolerance, as the accumulation of trap states due to radiation damage significantly affects the signal strength \cite{rad-damage}, a critical requirement for several physics experiments, as well as for medical \cite{medical} and space applications \cite{space}. \\
To overcome these limitations, several attempts are currently underway. A leading example is the successor of the ALPIDE chip being developed by the ALICE Collaboration for the ALICE ITS3 upgrade, where a modified process allows uniform depletion of the entire epitaxial layer, improving charge collection performance \cite{AGLIERIRINELLA_APTS}. \\
A further step has been taken by the ARCADIA project, led by an INFN collaboration \cite{da2025arcadia}, to collect charge from the entire device volume. By using a high-resistivity n-type substrate and deep p-wells in the epitaxial layer it is indeed possible to achieve complete and uniform depletion of the full sensing volume, realizing a Fully-Depleted MAPS (FD-MAPS). \\
This contribution details the early table-top characterization of the ARCADIA prototype chip. A quick overview of the chip is given in \autoref{sec:sensor}. Studies on threshold uniformity and calibration, done at the Physics and Astronomy Department of the University of Padova, are reported in \autoref{sec:threshold} and in \autoref{sec:energy_cal} respectively. \autoref{sec:charge_collection} focuses instead on charge collection efficiency measurements, performed with an infrared laser setup at Fermilab.

\section{Chip overview}
\label{sec:sensor}

The ARCADIA (Advanced Readout CMOS Architecture with Depleted Integrated sensor Arrays) sensor is a FD-MAPS developed by an INFN collaboration, and realized in a modified LFoundry 110nm CIS technology node on a high-resistivity substrate \cite{110nm-sensor}. Prototypes have been built with active thicknesses of 50, 100, and 200 \si{\micro\meter}.

\begin{figure}[ht]
    \centering
    \includegraphics[width=0.5\textwidth]{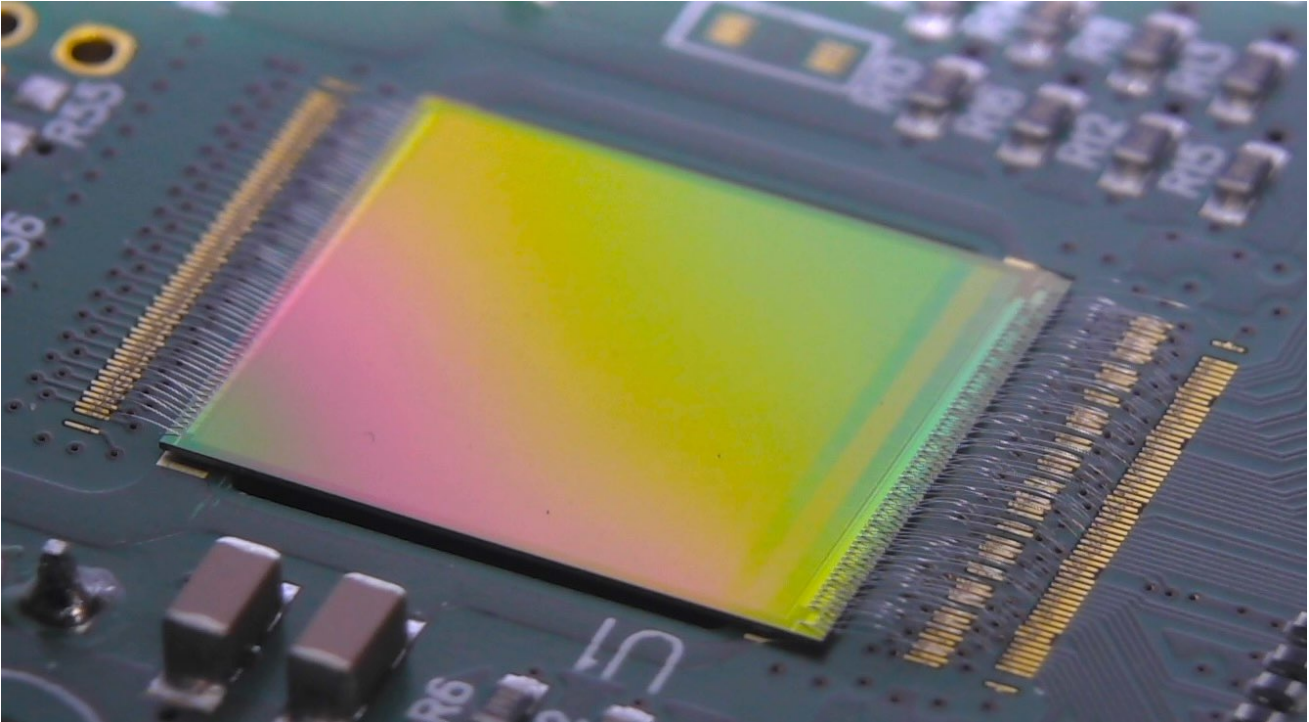}
    \caption{ARCADIA MD3 chip bonded on the PCB \cite{da2025arcadia}.}
    \label{fig:MD3_picture}
\end{figure}

\begin{figure}[ht]
    \centering
    \includegraphics[width=0.9\textwidth]{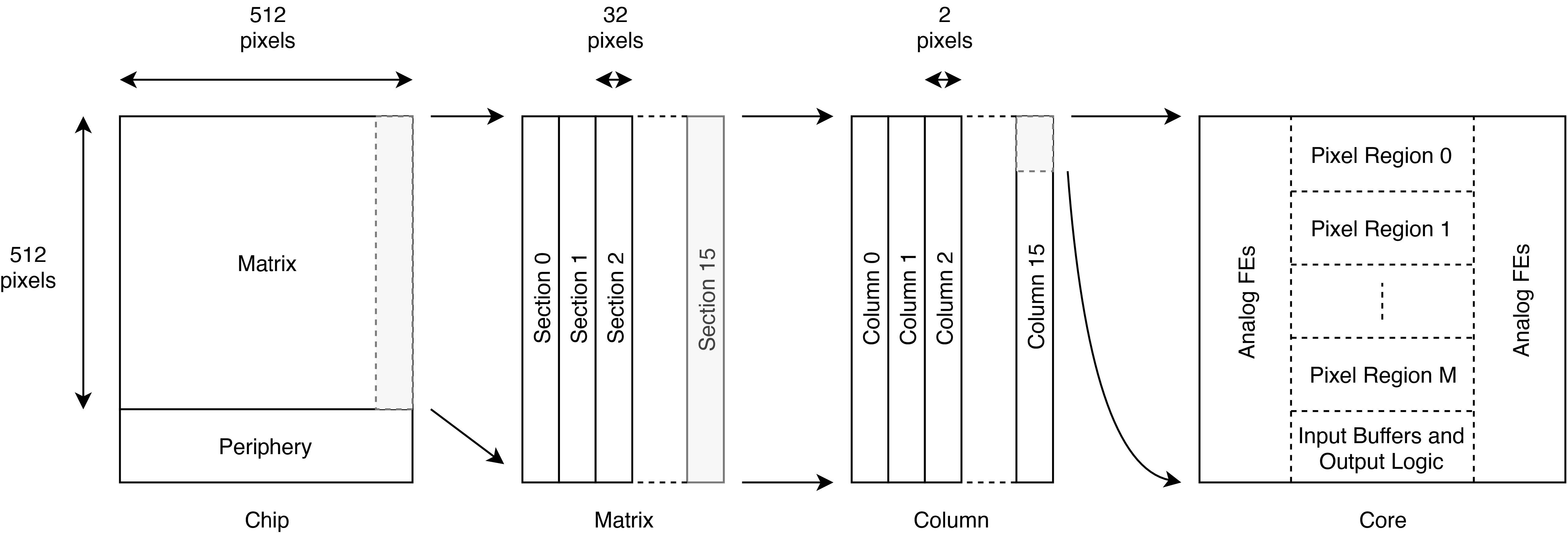}
    \caption{ARCADIA MD3 architecture organization \cite{da2025arcadia}.}
    \label{fig:MD3_architecture}
\end{figure}

This work focuses on the characterization of the first fully functional prototype, the so-called "Main Demonstrator 3" (MD3), shown in \autoref{fig:MD3_picture}. MD3 sports \SI{25}{\micro\meter} pitch square pixels arranged into a $512\times512$ array, for a total active area of $1.28\times1.28$ \SI{}{\square\centi\meter}. The pixel matrix is further divided into 16 independent sections of $512\times32$ pixels each.
In the pixel cell, a front-end amplifier senses the charge signal generated in the sensor. The preamplifier is connected to a simple comparator, which outputs a single bit, true whenever the amplifier output surpasses a given threshold. The threshold level is controlled through the VCASN register, connected to a 6-bit DAC, with bit resolution of \SI{5}{\milli\volt}. Each chip section has a separate DAC, allowing for per-section threshold tuning. While the actual threshold voltage is inversely proportional to the VCASN register decimal value, to ease data interpretation the threshold level is reported as $(63 - \text{VCASN})$ in all plots, to make it directly proportional to the actual voltage threshold, and expressed in DAC units. For testing purposes, pixels can be individually addressed through a tunable charge-injection circuitry.
Pixel cells are organized in 16 double columns of $512\times2$ pixels to form a section. Each double column forms an autonomous readout block, where pixels are arranged into regions of $4\times2$ pixels, and each pixel region connects to a shared control and data bus, as shown in \autoref{fig:MD3_architecture}. A combinatorial look-ahead token sequentially grants access to the data bus to any pixel region with a hit; when a pixel region accesses the bus, it downloads its address and the pixel hit-map to the periphery.

The periphery at the bottom of the pixel array implements all the control logic, together with the analog circuitry and several DACs sourcing the biasing currents and voltages. Like for the threshold one, all current and voltage DACs are replicated per chip section, to help fine tuning the operational working points across the sensor.
16 independent \qtyrange[range-units=single,range-phrase=-]{100}{320}{\mega\hertz} DDR serializers, one for each chip section, stream the data outside the sensor. For low data rate applications, e.g. space, it is possible to route all the data from the $16$ chip sections through a single serializer to enhance power efficiency. More details on the MD3 architecture are discussed in \cite{PhD_thesis_Davide, da2025arcadia}.
The MD3 device, object of this contribution, has a \SI{200}{\micro\meter} thick substrate that reach full depletion with a backside bias of \SI{-90}{\volt}. 

\section{Threshold uniformity}
\label{sec:threshold}

Threshold uniformity studies are performed both with test pulse injection circuitry and with a $^{55}$Fe radioactive source, as described respectively in \autoref{subsec:tp_scan} and \autoref{subsec:Fe55_scan}. \\
Since the performance of the front-end circuit is temperature dependent, the measurements are carried out in a temperature-controlled environment using an RTE-4DD refrigerated bath circulator with dry air flow. The temperature is monitored with a probe placed on the PCB and kept in the 20-30 \si{\degreeCelsius} range with a maximum variation of 2.5 \si{\degreeCelsius} during a single scan. \\
The rightmost chip section of the device tested in Padova does not work properly, which leads to only $512\times480$ testable pixels. Hence, all the data considered for the analysis and reported in the plots of \autoref{sec:threshold} and \autoref{sec:energy_cal} are cut at column 480.

\subsection{Test pulse scan}
\label{subsec:tp_scan}

\begin{figure}[ht]
    \centering
    \includegraphics[width=0.6\textwidth]{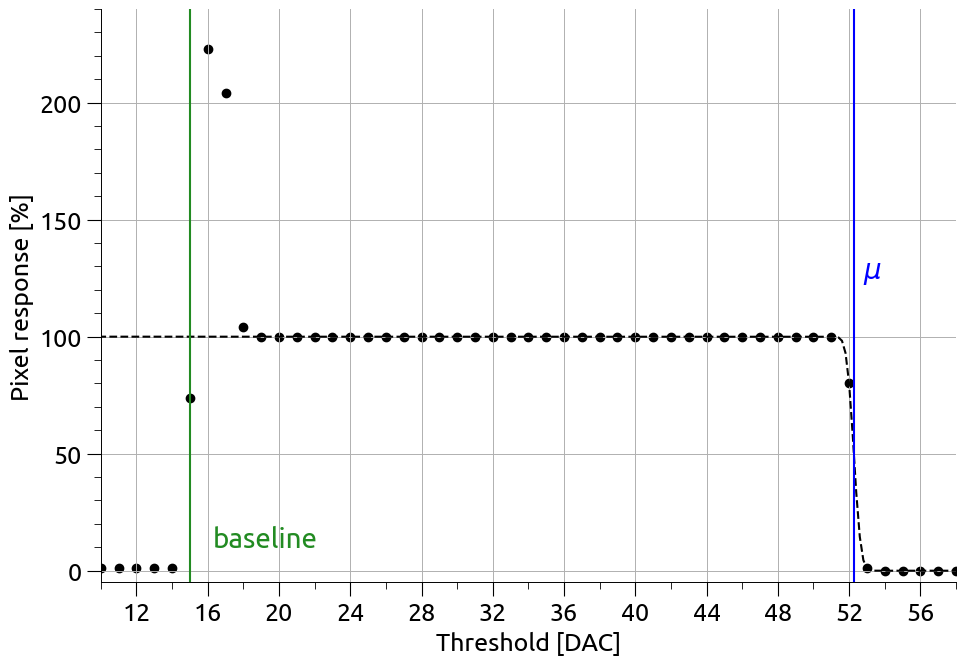}
    \caption{Single pixel S-curve fit: for threshold > $\upmu$, the pixel never responds to test pulse injection; for threshold around $\upmu$, pixel response rate starts to increase up to a region where the rate is always 100\%. 
    The noise region, i.e. where the pixel response rate is above 100\%, is entered when the threshold crosses the preamplifier baseline, causing the discriminator to trigger on noise. For even lower thresholds, the discriminator is always high and the pixel does not respond to any injection.}
    \label{fig:example_scurve_fit}
\end{figure}

Similar to other MAPS, ARCADIA MD3 pixels include a test-pulse injection circuit coupled to the pixel front end. The injected charge is loaded into a small capacitance (typically of the order of \SI{1}{\femto\farad}), and the amount of charge to be injected can be adjusted by setting two parameters. The small value of the injection capacitance is intentionally chosen to minimize the load on the input node, which represents a key advantage in terms of noise performance of the pixel. 
Because the capacitance is so small, parasitic effects can significantly alter the actual amount of injected charge, as discussed, for example, in \cite{Emiliani:2024zca}. Consequently, an absolute calibration of the front end using only test pulse injection is not reliable, and only a relative calibration can be achieved. This is further discussed in \autoref{sec:energy_cal}. \\
Threshold scans with test pulses are performed on single pixels to evaluate the uniformity of the threshold and baseline. In this procedure, a fixed number of injections is sent to a single pixel, and the number of responses is recorded for each threshold value $t$ in the scanned range. The pixel response can be described by the sigmoidal detector response S-curve given in \autoref{eq:TP_scurve}, where the midpoint $\upmu$ corresponds to the fitted value at 50\% hit probability. An example of a S-curve fit is shown in \autoref{fig:example_scurve_fit}.

\begin{equation}
    \mathrm{R(t) = \frac{1}{2} \left[1+erf\left(\frac{t-\mu}{\sqrt{2}\sigma}\right)\right]}
    \label{eq:TP_scurve}
\end{equation}

\begin{figure}[ht]
    \centering
    \begin{subfigure}{0.48\textwidth}
    \includegraphics[height=5.5cm]{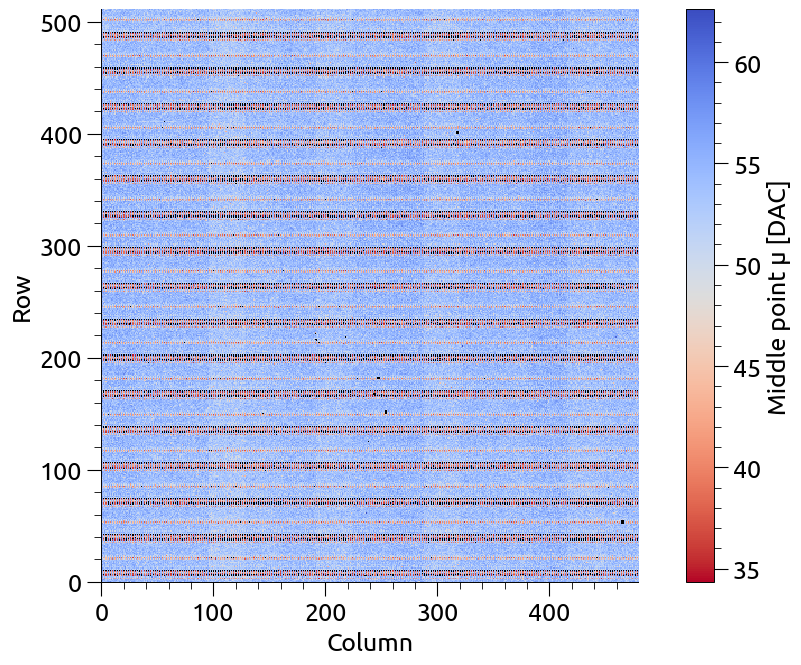}
    \caption{}     
    \label{fig:heatmap_TP}    
    \end{subfigure}
    \hfill
    \begin{subfigure}{0.5\textwidth}
    \includegraphics[height=5.5cm]{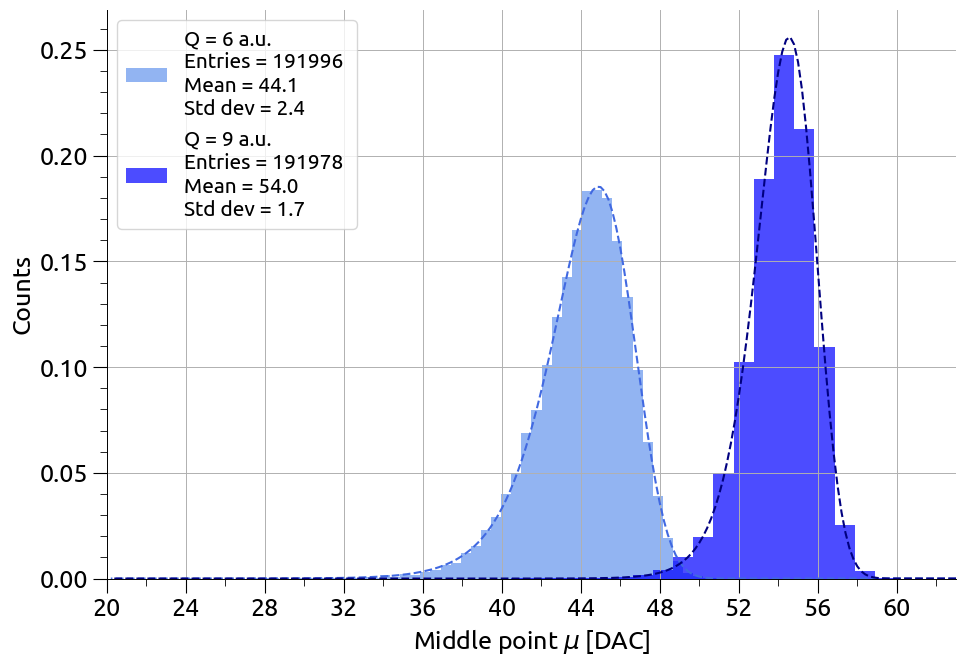}
    \caption{}
    \label{fig:thr_distr_TP}    
    \end{subfigure}
    \caption{Pixels S-curve middle point map \subref{fig:heatmap_TP} of threshold scan with 9 a.u. injected charge. Black entries are pixels that do not show a S-curve behavior in the scan range of [33, 63] DAC, hence lacking any actual middle point. Distribution of middle point values \subref{fig:thr_distr_TP} for the subsample of pixels not affected by the injection issue, for 6 a.u. and 9 a.u. charge injected.
    }
\end{figure}

\begin{figure}[ht]
    \centering
    \begin{subfigure}{0.5\textwidth}
    \includegraphics[height=5.5cm]{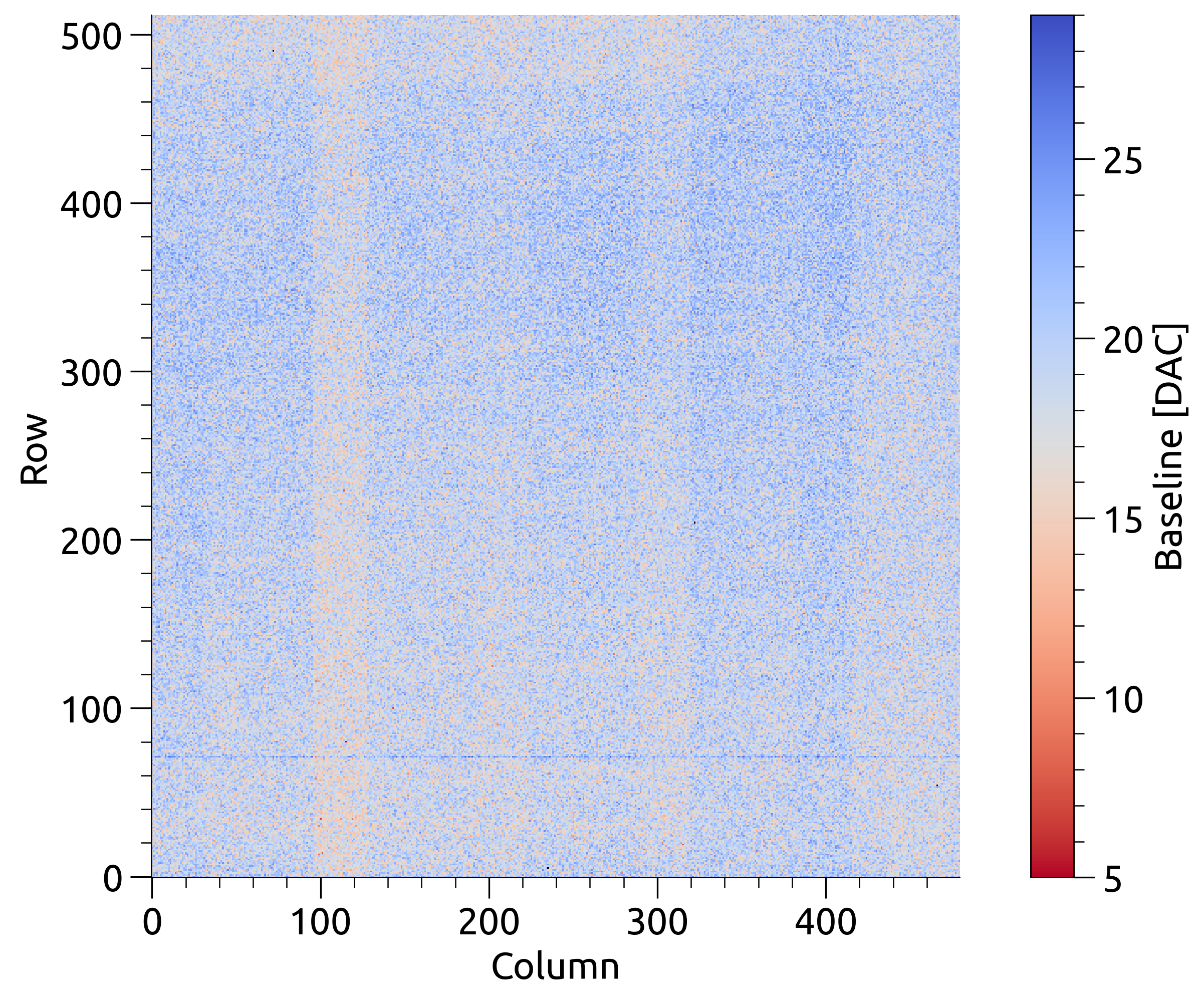}
    \caption{}
    \label{fig:heatmap_baseline}    
    \end{subfigure}
    \hfill
    \begin{subfigure}{0.48\textwidth}
    \includegraphics[height=5.5cm]{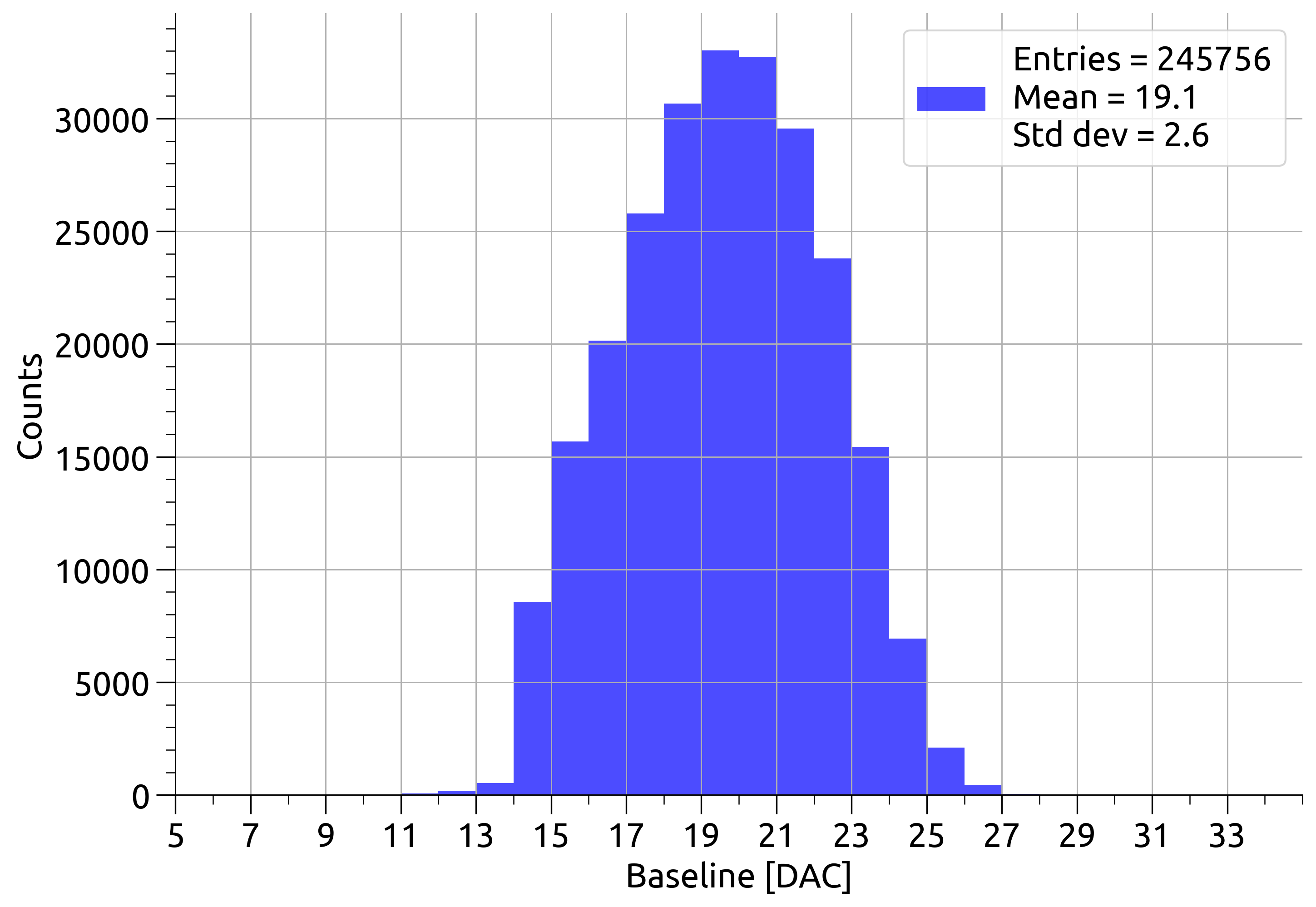}
    \caption{}
    \label{fig:baseline_distr_TP}
    \end{subfigure}
    \caption{Baseline map \subref{fig:heatmap_baseline} and baseline distribution \subref{fig:baseline_distr_TP} obtained from test pulse measurements.}
    \label{fig:baseline_distributions}
\end{figure}

The $\upmu$ value for each pixel is obtained from the single-pixel S-curve fit. As shown in \autoref{fig:heatmap_TP}, a visible pattern along the rows emerges, with certain pixels consistently showing a significantly lower $\upmu$ for the same injected charge. 
The pattern is not observed in threshold scans with physical sources (see \autoref{fig:thr_map_Fe55}); therefore, it is attributed to capacitive coupling effects that modify the injection capacitance at specific pixel positions. Further investigation of this effect is ongoing. Taking into account only pixels unaffected by this issue (78\% of the testable pixels), the distribution of $\upmu$ is shown in \autoref{fig:thr_distr_TP} for two different values of injected charge. To obtain the conversion from the injected charge in $\text{a.u.}$ to electrons \autoref{conversion} can be used. Since the response of the front-end amplifier is not linear, the distributions of \autoref{fig:thr_distr_TP} are fitted with the convolution of a Gaussian with a lognormal function. The standard deviation of the histogram gives a relative spread of 5.4\% for the lowest and 3.1\% for the highest injected charge value. \\
From \autoref{fig:thr_distr_TP} it is also possible to observe that some chip sections exhibit a lower $\upmu$ value. The same effect is also visible in \autoref{fig:heatmap_baseline} and \autoref{fig:thr_map_Fe55}. If needed, the threshold value can be properly tuned section by section in order to obtain a more uniform chip response. 

\autoref{fig:baseline_distributions} shows the distributions of the pixel baseline, defined as the threshold value below which the output of the pixel comparator stays high. In this work the baseline is estimated as the first threshold value at which the number of test-pulse responses falls below the number of injections, as shown in the S-curve example in \autoref{fig:example_scurve_fit}. The baseline distribution in \autoref{fig:baseline_distr_TP} is peaked at a value of 19 DAC with a spread of less than 14\%.

\subsection{\texorpdfstring{$^{55}$Fe}{55Fe} scan}
\label{subsec:Fe55_scan}

A $^{55}$Fe source is also used to perform threshold studies measuring the hit rate as a function of threshold. The source is placed at a distance such that the illumination is uniform across the whole pixel array. In this case, the signal is generated by charges collected from the sensor substrate and, due to the contribution of diffusion, the resulting S-curves are described by a more complex function, given in \autoref{eq:scurve} \cite{scurve}.

\begin{equation}
    \mathrm{N(t) = N_0 \left[ 1 + \frac{C_s}{\sigma}(\mu - t) \right]\left[ 1 + erf\bigg( \frac{\mu - t}{\sqrt{2}\sigma}\bigg)\right]}
    \label{eq:scurve}
\end{equation}

\begin{figure}[ht]
    \centering
    \includegraphics[height=7cm]{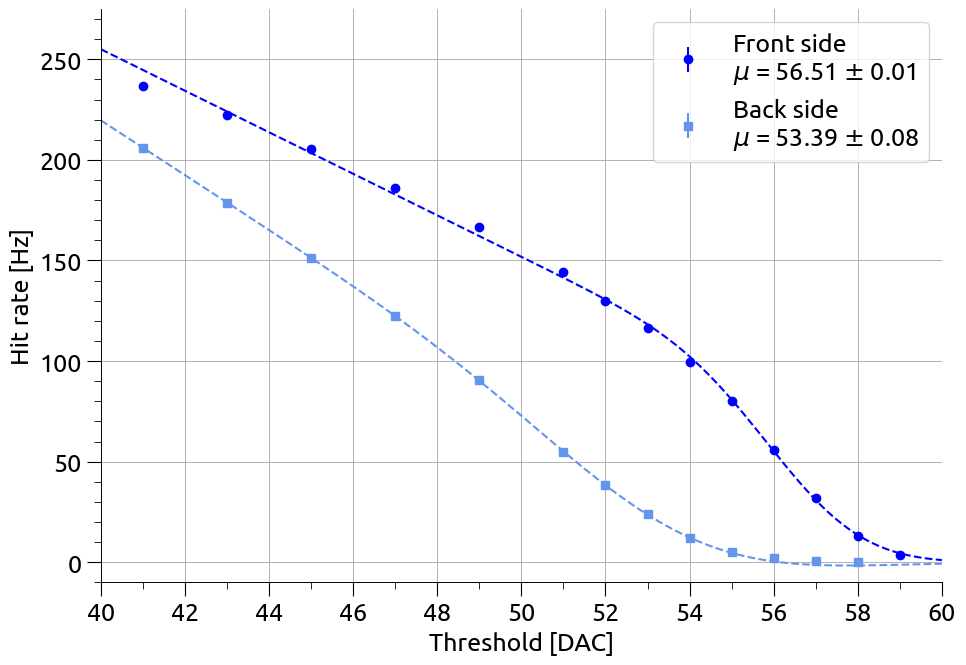}
    \caption{Comparison of the S-curve obtained on the $16\times16$ pixel array for front- and back-side illumination with $^{55}$Fe.}
    \label{fig:scurve_front_back}
\end{figure}

The second term of \autoref{eq:scurve} corresponds to the conventional S-curve function, while the first term models charge sharing (the C$_s$ term in the equation, normalized to $\upsigma$ to be dimensionless), which leads to a linear increase in the hit rate for threshold values below $\upmu$. \\
On a smaller sub-array of $16 \times 16$ pixels, fully contained within chip section 8, a threshold scan is performed with the source positioned on both the front (charge-collection side) and back sides. Being the chip substrate fully depleted, charges generated throughout the entire substrate thickness can be collected. At the $^{55}$Fe K$\upalpha_{1}$ peak energy (\SI{5.89}{\kilo\eV} \cite{mougeot2025evaluations}), the attenuation of X-rays in \SI{50}{\micro\meter} of silicon is about 83.4\% \cite{Xray_absorption}. Consequently, under backside illumination, charges are generated on average farther from the collection electrodes, resulting in greater diffusion and thus increasing charge sharing. A comparison of the S-curves obtained with front- and back-side illumination at the same temperature is shown in \autoref{fig:scurve_front_back}. From the slope of the linearity region (low threshold range), it can be observed that the diffusion is higher in case of backside illumination. As a consequence, the middle point $\upmu$ corresponding to the same number of generated electrons, is lower by 3 DAC ($\simeq$ 300 e$^-$, see \autoref{subsec:calibration}) units for the backside case.

\begin{figure}[ht]
    \centering
        \begin{subfigure}{0.49\textwidth}
        \centering
        \includegraphics[height=5.5cm]{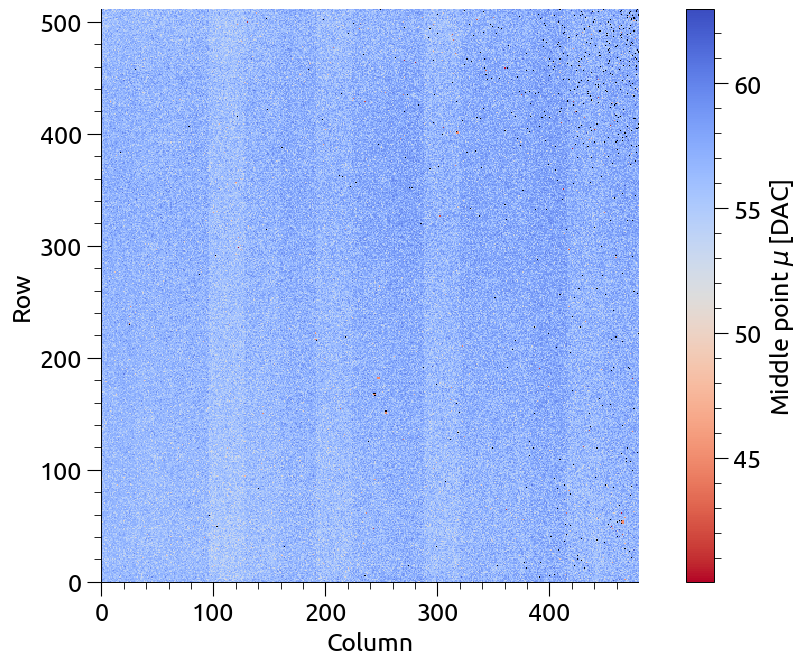}
        \caption{}
        \label{fig:thr_map_Fe55}
    \end{subfigure}
    \hfill
    \begin{subfigure}{0.49\textwidth}
        \centering
        \includegraphics[height=5.5cm]{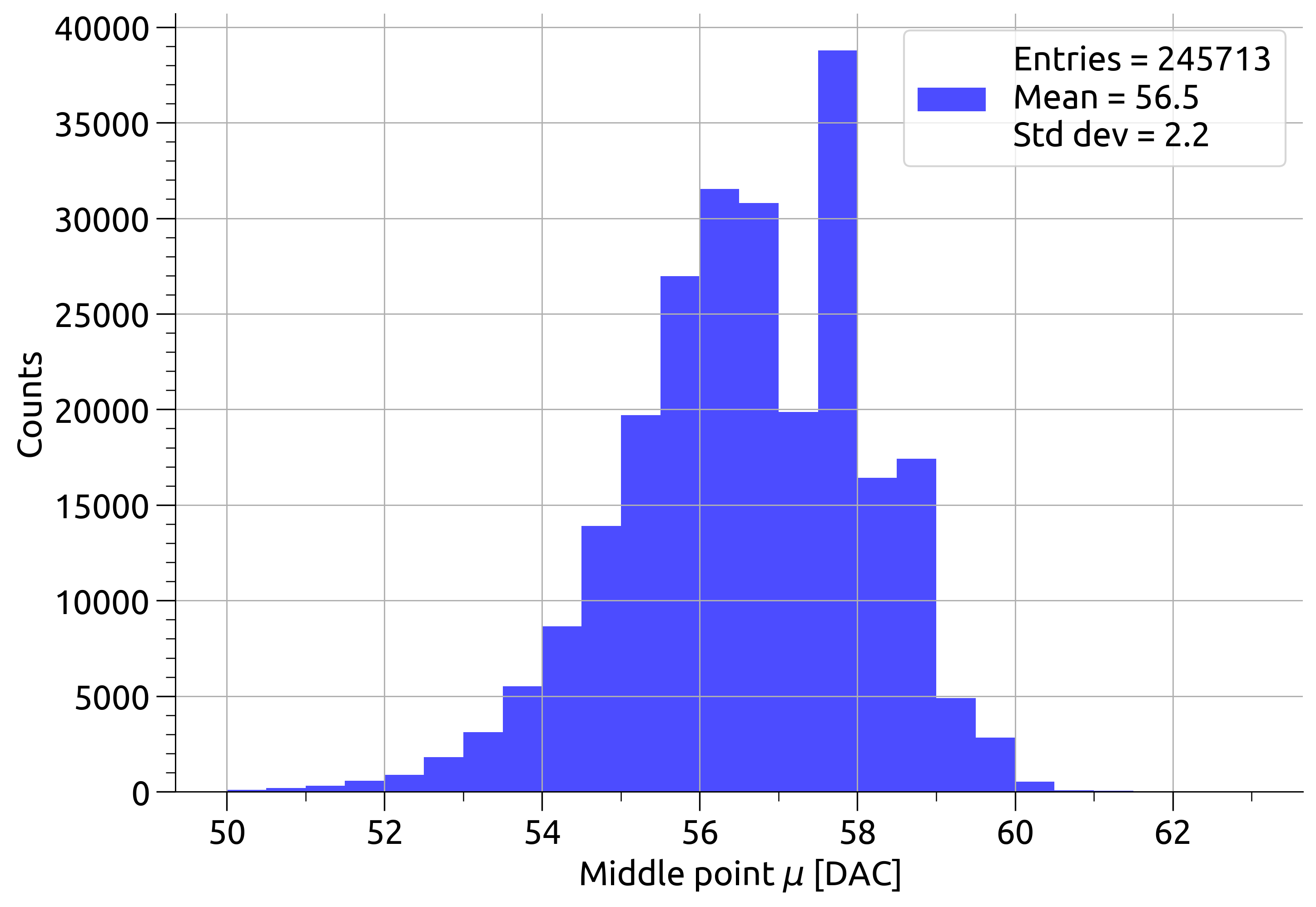}
        \caption{}
        \label{fig:thr_hist_Fe55}
    \end{subfigure}
    \caption{Pixels S-curve middle point map \subref{fig:thr_map_Fe55}: black entries represent pixels that do not show a S-curve behavior, therefore it is not possible to compute the corresponding middle point value. Distribution of middle point values \subref{fig:thr_hist_Fe55} of threshold scan measurements with $^{55}$Fe source.}
    \label{fig:thr_distr_Fe55}
\end{figure}

On the full pixel array, a threshold scan is performed with the source placed on the front side of the chip to study the uniformity of the response to ionizing radiation within the sensing volume.
The number of hits per pixel is recorded, and the rate as a function of threshold is fitted using \autoref{eq:scurve}. The distribution of single-pixel middle points $\upmu$, is shown in \autoref{fig:thr_hist_Fe55}. The mean of the distribution represents the threshold that corresponds to approximately 1640 electrons, and the relative spread of the distribution is of 3.9\%.

\section{Energy calibration}
\label{sec:energy_cal}

To perform an energy calibration, additional threshold-scan measurements are carried out using monochromatic fluorescence X-rays. This is a standard procedure for the energy calibration of digital sensors with an analogue output, and is typically performed using Time-over-Threshold (ToT) measurements \cite{rinella2023digital, AGLIERIRINELLA2026171082}. Since the MD3 device has no analogue output, the fluorescence X-ray characterization is performed by measuring the hit rate as a function of the threshold, as already done for the characterization of other digital MAPS \cite{Malta2}. The goal is to obtain S-curves corresponding to different characteristic fluorescence X-ray energies, allowing a conversion from DAC units to charge.

\subsection{Fluorescence X-rays}
\label{subsec:fluorescence}

\begin{figure}[ht]
    \centering
    \begin{subfigure}[t]{0.4\textwidth}
        \centering
        \includegraphics[height=5cm]{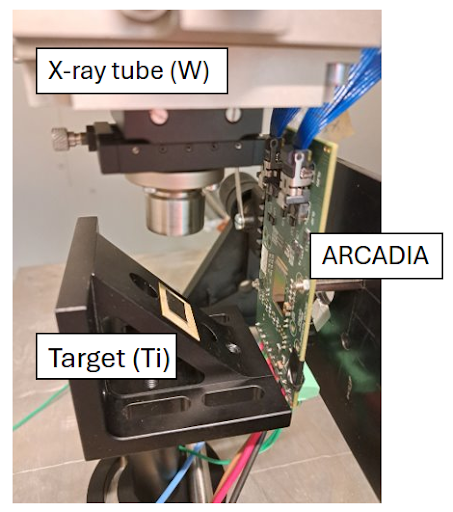}
        \caption{}
        \label{fig:fluo_setup}
    \end{subfigure}
    \hfill
    \begin{subfigure}[t]{0.55\textwidth}
        \centering
        \includegraphics[height=5cm]{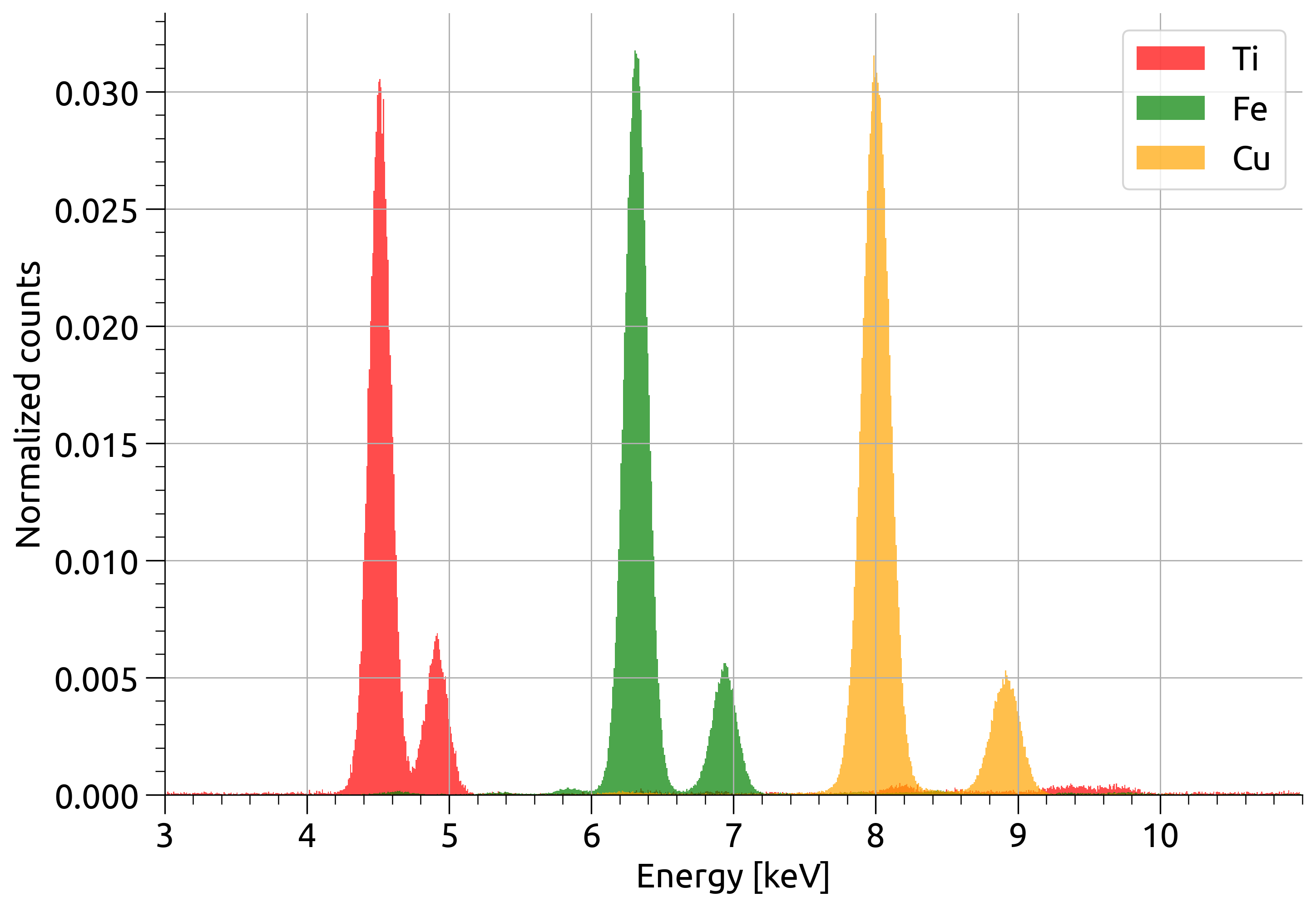}
        \caption{}
        \label{fig:fluo_Ti_spectrum}
    \end{subfigure}
    \caption{Experimental setup \subref{fig:fluo_setup} for fluorescence X-ray measurements and fluorescence spectra of the three targets \subref{fig:fluo_Ti_spectrum} measured with the AMPTEK XR-100CR Si-PIN diode detector \cite{Amptek}.}
    \label{fig:fluo}
\end{figure}

The experimental setup for fluorescence measurements, shown in \autoref{fig:fluo_setup}, consists of an X-ray machine and targets made of different materials. The machine is a Seifert Model RP149 equipped with a tungsten-anode X-ray tube, with bias voltage and anode current settings of up to \SI{40}{\kilo\volt} and \SI{50}{\milli\ampere}, respectively. A collimator is used to narrow the beam. To suppress the primary X-ray component, the target is placed below the X-ray tube at an angle of \SI{45}{\degree} with respect to the primary beam axis, while the sensor under test is positioned vertically with the front-side of the chip facing the target. Thermal control is not possible with this X-ray tube setup; however, the temperature is monitored during the measurements. A probe placed on the PCB records an average temperature of \SI{28}{\degreeCelsius}, with a maximum variation of \SI{1}{\degreeCelsius} during a complete threshold scan.

\begin{figure}[ht]
    \centering
    \begin{subfigure}[t]{0.49\textwidth}
    \includegraphics[height=5cm]{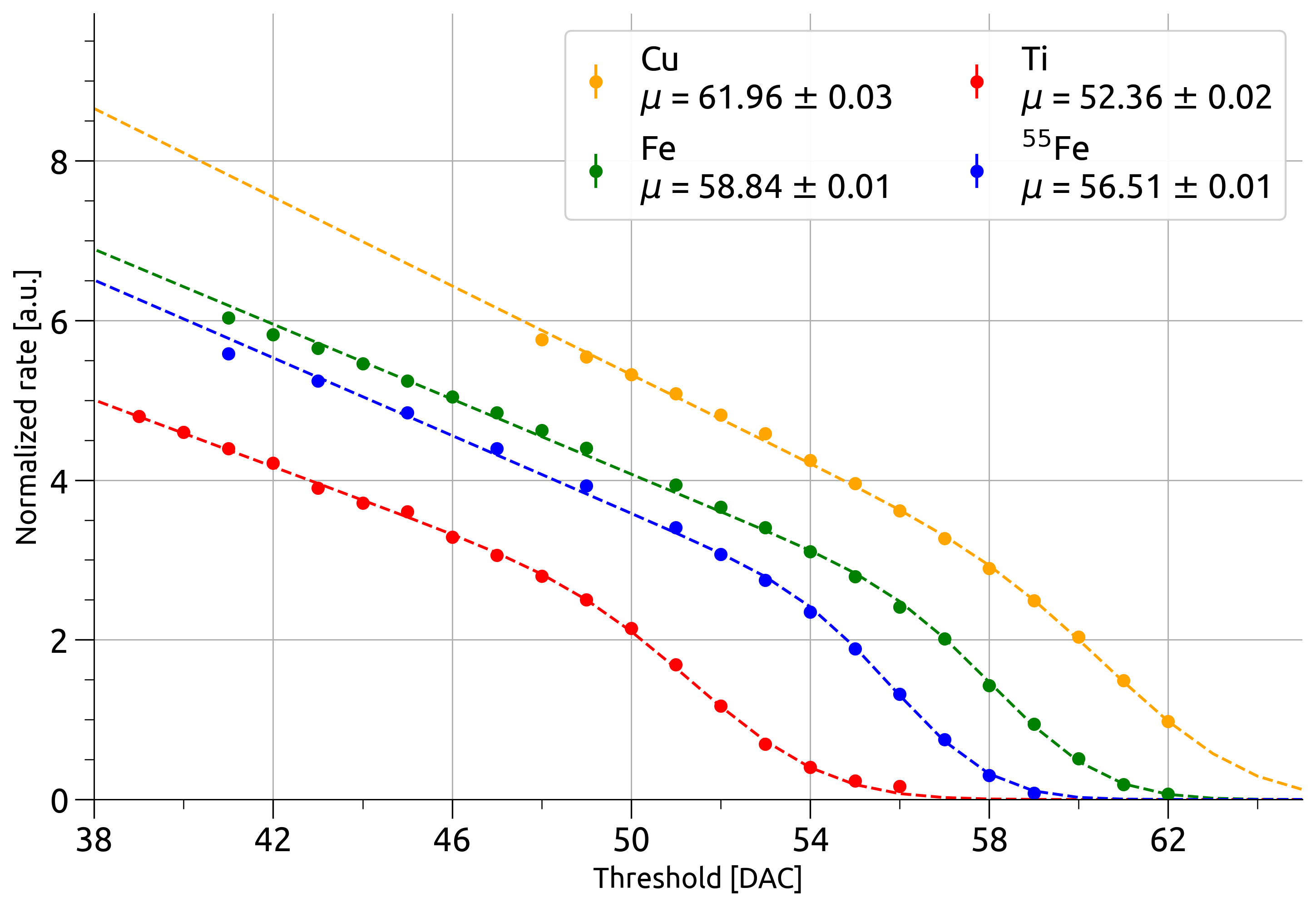}
    \caption{}
    \label{fig:scurve_xrays_full_array}    
    \end{subfigure}
    \begin{subfigure}[t]{0.49\textwidth}
        \includegraphics[height=5cm]{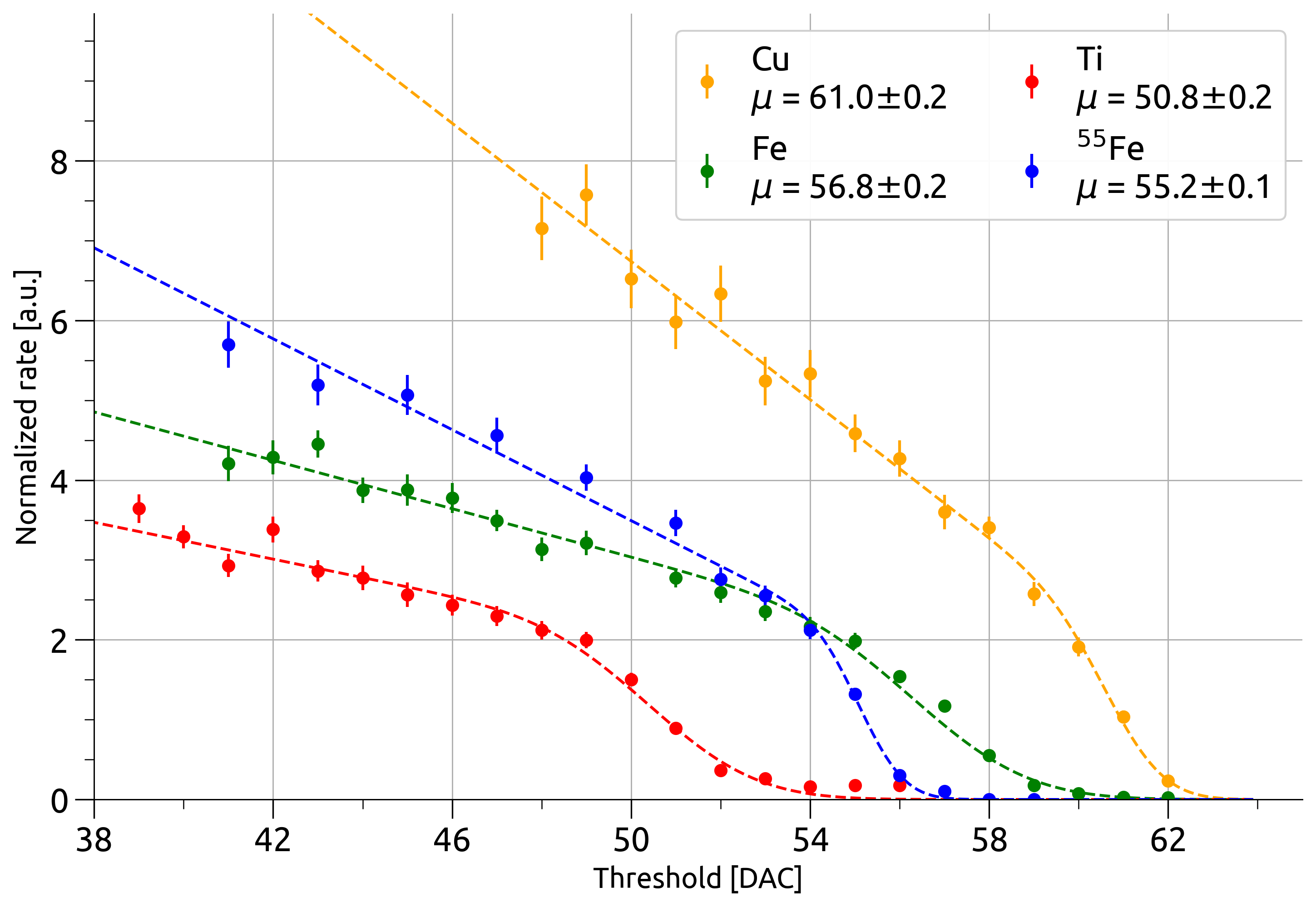}
        \caption{}
        \label{fig:scurve_xrays_single_pix}
    \end{subfigure}
    \caption{S-curve comparison for X-ray fluorescence (Ti, Fe and Cu targets) and $^{55}$Fe measurements with the sensor illuminated from the front side: \subref{fig:scurve_xrays_full_array} for the full $16\times16$ sub-array and \subref{fig:scurve_xrays_single_pix} for one single pixel}
    \label{fig:scurve_xrays}
\end{figure}

Measurements are performed using three targets: titanium, iron, and copper. According to the X-ray Data Booklet \cite{xraydatabooklet2009}, their K$\upalpha_{1}$ peak energies are 4.51, 6.40, and 8.08 \si{\kilo\eV}, respectively (\autoref{fig:fluo_Ti_spectrum}). For each target, a threshold scan is performed on the same $16\times16$ pixel sub-array used for the $^{55}$Fe front- and back-side measurements.\\ The results are shown in \autoref{fig:scurve_xrays}. The plot in \autoref{fig:scurve_xrays_full_array} shows the hit rate on the full sub-array as a function of the threshold. The hit rate is normalized such that the middle point corresponds to a unitary rate. As expected, the S-curves move towards lower threshold as the X-rays characteristic energy decreases.

\subsection{Threshold and noise calibration}
\label{subsec:calibration}

As mentioned in \autoref{sec:threshold}, threshold scans with test pulses can provide only a relative calibration of the pixel threshold. The characterization of the $16\times16$ pixel array with monochromatic X-rays can provide instead an absolute calibration of the test pulse injected charge. With the procedure described below, a relation between the injected charge in $\text{a.u.}$ and the number of electrons released by monochromatic X-rays is established. The procedure is done for every single pixel of the sub-array.

\begin{figure}[ht]
    \centering
    \begin{subfigure}[t]{0.48\textwidth}
        \centering
        \includegraphics[height=5cm]{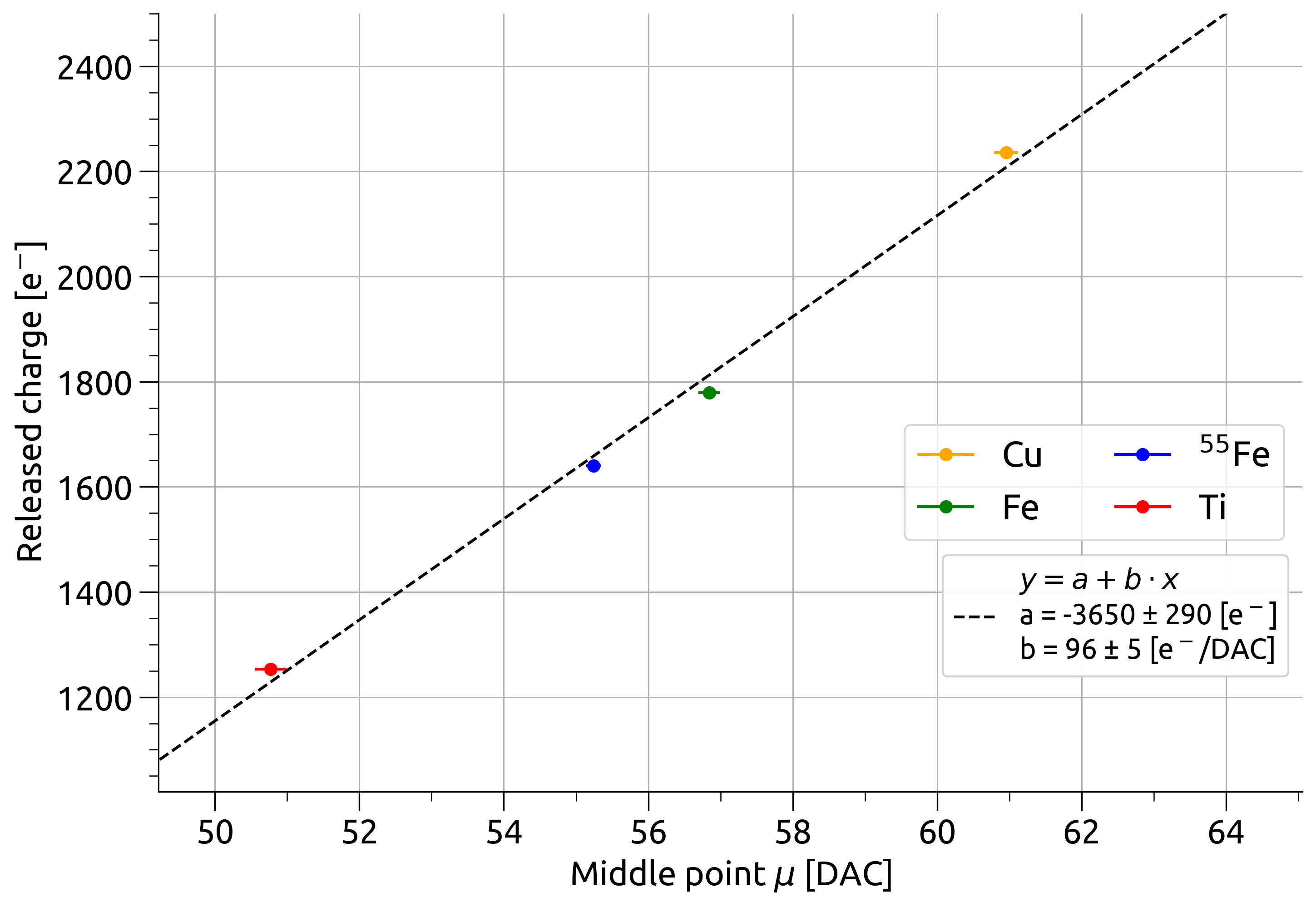}
        \caption{}
        \label{fig:lin_fit_345_275}
    \end{subfigure}
    \begin{subfigure}[t]{0.48\textwidth}
        \centering
        \includegraphics[height=5cm]{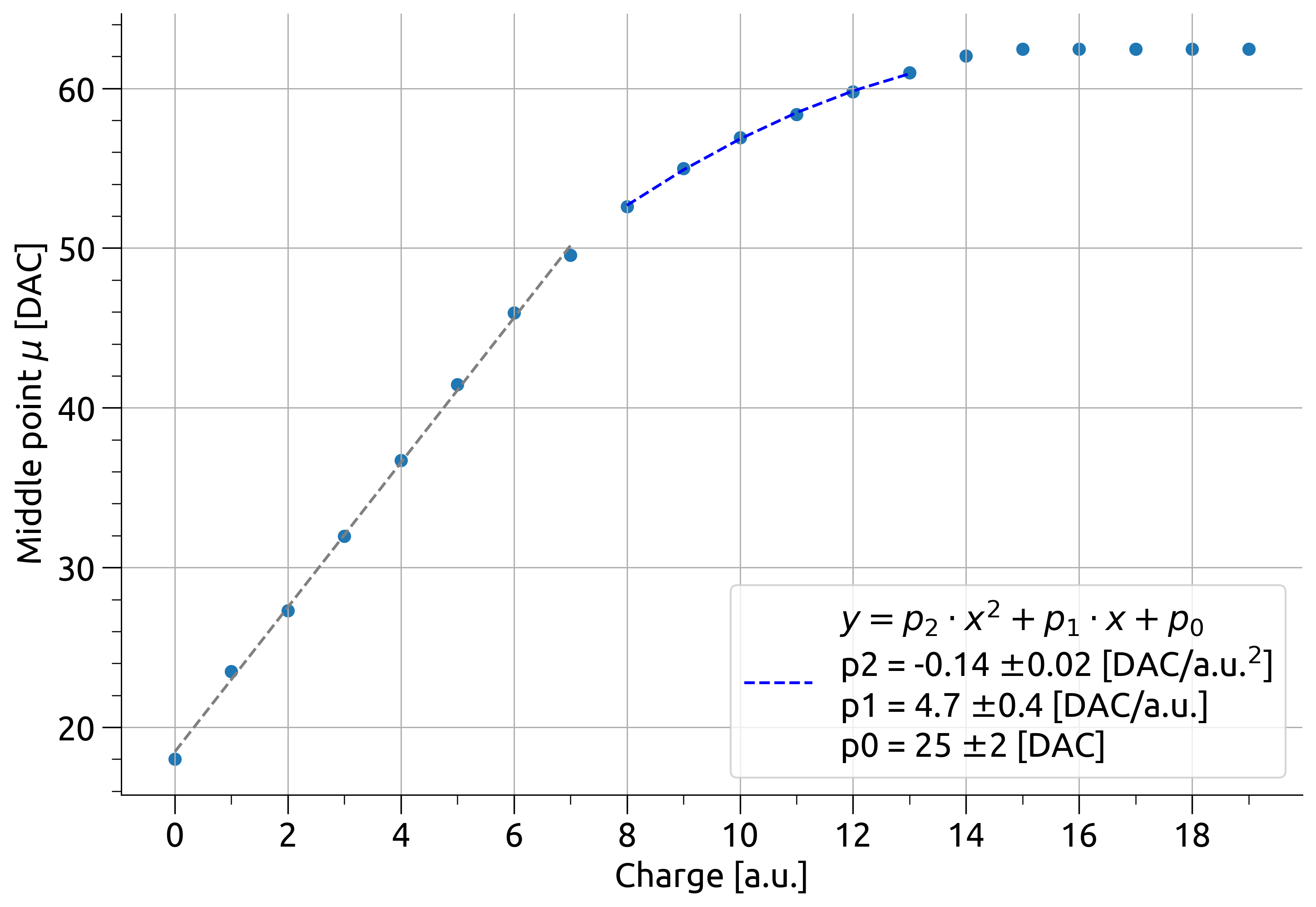}
        \caption{}
        \label{fig:poly_fit_345_275}
    \end{subfigure}
    \caption{Linear fit of X-ray released charge versus S-curve middle point \subref{fig:lin_fit_345_275} and S-curve middle point in DAC units as a function of the injected charge \subref{fig:poly_fit_345_275} for pixel (345,275).}
    \label{fig:calibration_single_pixel}
\end{figure}

\begin{enumerate}[label=\text{\arabic*)}]
    \item As a first step, the single-pixel S-curves are obtained for the four X-ray characteristic energies as shown in \autoref{fig:scurve_xrays_single_pix}. The relation between the number of electrons and DAC units is determined by a linear fit of the X-ray K$\upalpha_{1}$ peak energy as a function of the corresponding S-curve middle point. An example for one pixel, together with the fit result, is shown in \autoref{fig:lin_fit_345_275}. \label{first-item}
    
    \item A relation between DAC units and the arbitrary units of injected charge is obtained from test pulse threshold scans. \autoref{fig:poly_fit_345_275} shows the S-curve midpoint in DAC units as a function of the injected charge. Two distinct regions are observed: a linear region at low injected charge, and a saturation region at high injected charge. In the range explored with the X-ray measurements (above 50 DAC), the dependence of the threshold on the injected charge is well described by a second-degree polynomial, which can be used to convert arbitrary injected charge values into DAC units. \label{second-item}
    
    \item By combining the relations obtained in points \ref{first-item} and \ref{second-item}, a calibration curve can be derived to convert the injected charge $q_{\mathrm{a.u.}}$ (arbitrary units) into electrons $q_{e^{-}}$. 
    
    \begin{equation}
    \label{conversion}
        \text{q$_{e^-}$} = \text{a} + \text{b} (\text{p}_2 \cdot \text{q}^2_{\text{a.u.}} + \text{p}_1 \cdot \text{q}_{\text{a.u.}}  + \text{p}_0) 
    \end{equation}

\end{enumerate}

\begin{figure}[ht]
    \centering
    \begin{subfigure}[t]{0.45\textwidth}
        \centering
        \includegraphics[width=\linewidth]{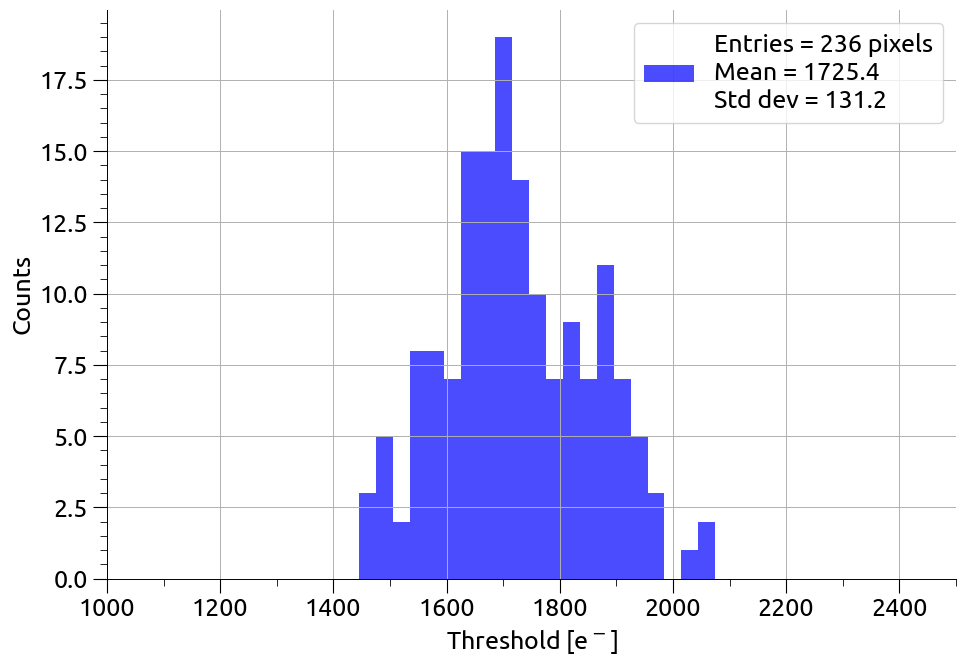}
        \caption{}
        \label{fig:threshold_142_pixels_electrons}
    \end{subfigure}
    \hfill
    \begin{subfigure}[t]{0.45\textwidth}
        \centering
        \includegraphics[width=\linewidth]{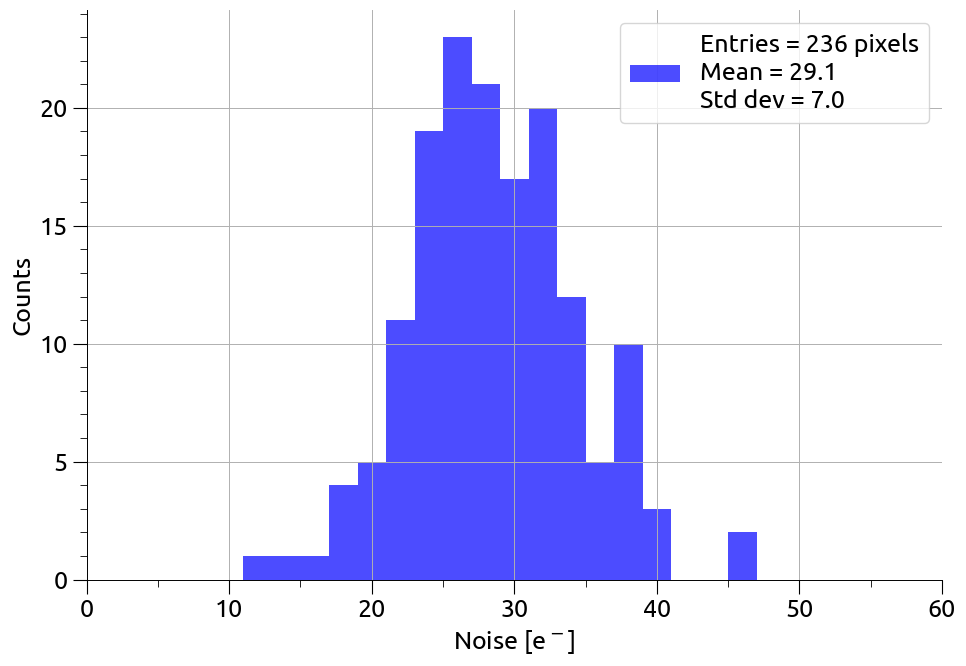}
        \caption{}
        \label{fig:noise_142_pixels_electrons}
    \end{subfigure}
    \caption{Threshold \subref{fig:threshold_142_pixels_electrons}  and noise \subref{fig:noise_142_pixels_electrons} distributions in electrons.}
    \label{fig:thr_noise_calib}
\end{figure}

Threshold and noise in electrons are estimated through the previous procedure for the test pulse injected charge at the threshold value of 58 DAC, assumed as reference for the noise calibration. The two values are extracted from the fit (\autoref{eq:TP_scurve}) of the calibrated S-curve in the [50, 62] DAC range. \autoref{fig:thr_noise_calib} shows the threshold and noise distributions expressed in terms of electrons. Only pixels satisfying two conditions are included in the final distributions: valid DAC value for at least three out of four X-ray S-curve fits for step \ref{first-item}, and at least one point in the central part of the S-curve of the injected charge scan to ensure a reliable noise estimation for step \ref{second-item}. \\
For the 236 pixels satisfying the previous conditions, the average number of electrons corresponding to a threshold of 58 DAC is 1725 electrons. The threshold distribution (\autoref{fig:threshold_142_pixels_electrons}) has a standard deviation that corresponds to a relative width of 7.6\%. The average noise at the same threshold value is 29 electrons, with a large relative spread of 24\% (\autoref{fig:noise_142_pixels_electrons}).

\section{Charge collection efficiency}
\label{sec:charge_collection}

Selected areas of the pixel array are scanned to characterize the charge collection efficiency using an IR laser. For reference, ARCADIA passive test structures had previously been measured using an IR and a visible red laser \cite{laser_paper}.

\subsection{Experimental Setup}
\label{sec:laser}
\begin{figure}[ht]
    \centering
    \begin{subfigure}{0.49\textwidth}
    \includegraphics[height=5.5cm]{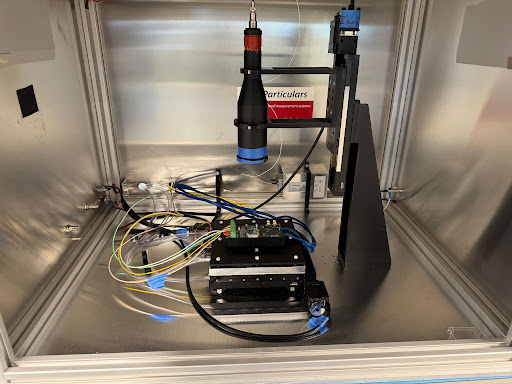}
    \caption{}
    \label{fig:laser_setup}
    \end{subfigure}
    \begin{subfigure}{0.49\textwidth}
    \includegraphics[height=5.5cm]{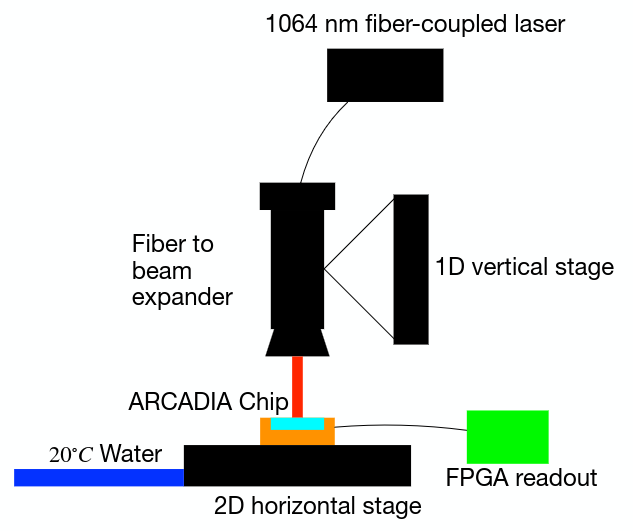}
    \caption{}
    \label{fig:scan_central.laser}
    \end{subfigure}
    \caption{
        An overview of the laser testing apparatus with \subref{fig:laser_setup} the shielded box containing an IR laser, XYZ linear stage, and the ARCADIA chip, and \subref{fig:scan_central.laser} a sketch of the apparatus with labels corresponding to the parts in \subref{fig:laser_setup}. The FPGA readout board is not seen as it is placed outside of the box.
    }
    \label{fig:laser.apparatus}
\end{figure}

The apparatus sits inside a box with metal to suppress electromagnetic noise and ambient light, as shown in \autoref{fig:laser.apparatus}. A fiber-coupled \SI{1}{\kilo\hertz} pulsed laser with a wavelength of \SI{1064}{\nano\meter} is connected to a beam expander and a manual iris shutter to produce a Gaussian beam.

Two linear-axis stages move the mounted sensor perpendicular to the laser with a step size precision of less than \SI{1}{\micro\meter}, while a third stage provides vertical motion to focus the laser spot to an estimated full width at half maximum (FWHM) of less than \SI{10}{\micro\meter}.
The sensor is illuminated through the backside, with a hole cut in the PCB to expose the entire chip. An external water cooler at \SI{20}{\degreeCelsius} maintains a stable operating temperature. The starting pixel is chosen near the center of the chip, and aligned with the laser to a precision of \SI{0.5}{\micro\meter}. Each scanning step accounts for the rotation between the sensor and the 2D linear-axis stages.

A region of $10\times6$ pixels at the center of the chip is characterized. During a scan, each pixel is covered by a grid pattern with a step size of \SI{5}{\micro\meter}, corresponding to 25 scans per pixel as  shown in \autoref{fig:laser.inpixel_scheme}.
At each position, the threshold is varied from 48 to 62~DAC in steps of 2. For every threshold value, the total number of hits is recorded and the efficiency is measured.
The efficiency in \autoref{eqn:eff_hits} is defined as the ratio between the number of recorded hits, and the total number of laser pulses, computed as the product of the integration time window and the measured laser frequency. 

\begin{equation}
    \varepsilon = \frac{\mathrm{hits}}{t \cdot f}
    \label{eqn:eff_hits}
\end{equation}

\subsection{Results}

\begin{figure}[ht]
    \centering
    \begin{subfigure}{0.3\textwidth}
    \centering
    \includegraphics[width=\linewidth]{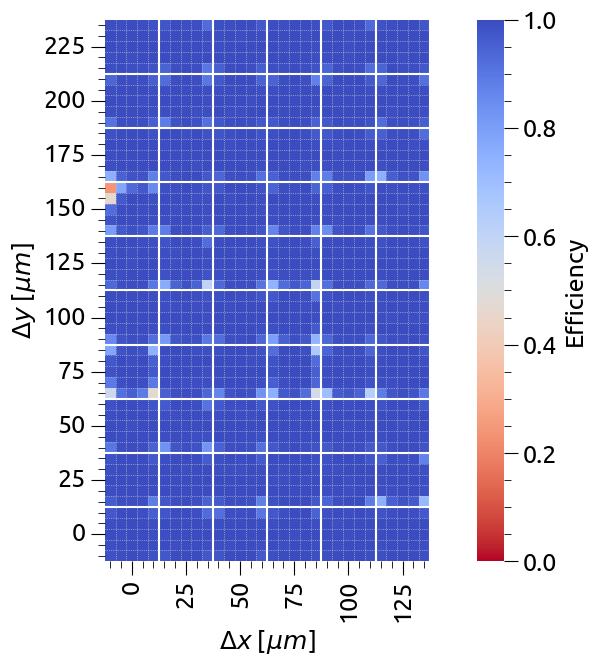}
    \caption{}
    \label{fig:laser-eff}
    \end{subfigure}
    \begin{subfigure}{0.3\textwidth}
    \centering
    \includegraphics[width=\linewidth]{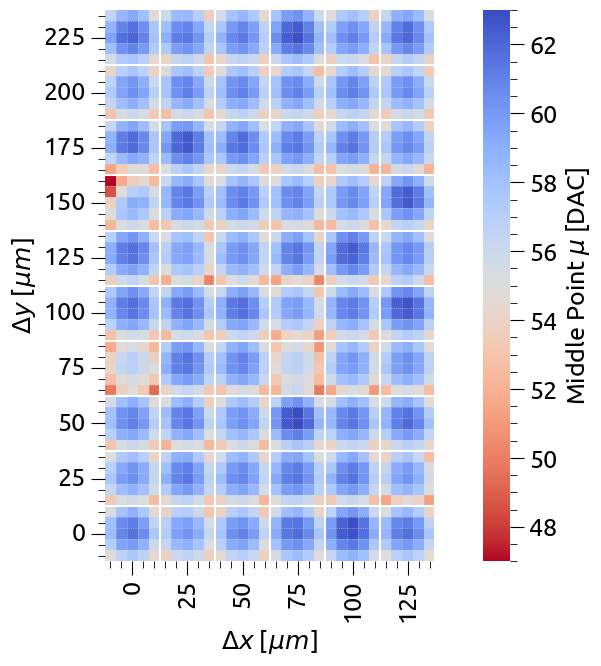}
    \caption{}
    \label{fig:laser-middle-point}
    \end{subfigure}
    \begin{subfigure}{0.3\textwidth}
    \centering
    \includegraphics[width=\linewidth]{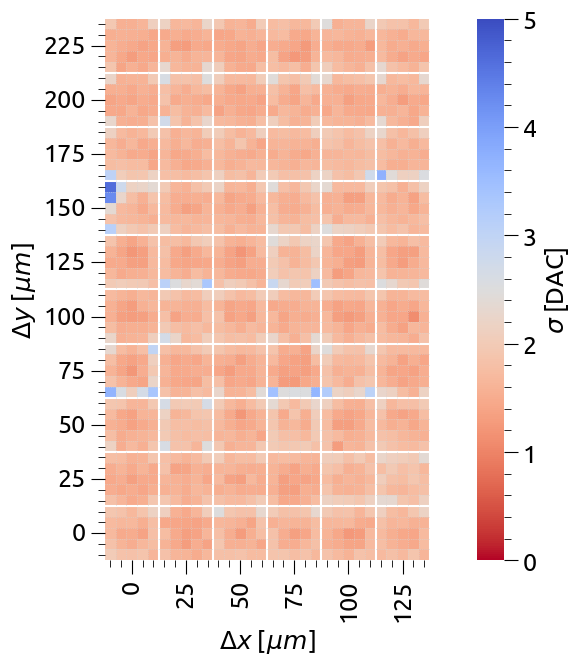}
    \caption{}
    \label{fig:laser-sigma}
    \end{subfigure}
    \caption{
        The measurements of the efficiency at \subref{fig:laser-eff} threshold 52 DAC, \subref{fig:laser-middle-point} the middle point $\upmu$, and \subref{fig:laser-sigma} $\upsigma$ of a $10\times6$ pixel sub-array. The $\Delta x$ and $\Delta y$ are relative positions from the center of the first scanned pixel at the bottom left.}
    \label{fig:laser.cluster}
\end{figure}

The efficiency as a function of the threshold is fitted with an S-curve function, as defined in \autoref{eq:TP_scurve}. The resulting $\upmu$ and $\upsigma$ are shown in \autoref{fig:laser-middle-point} and \autoref{fig:laser-sigma}. The efficiency map at 52 DAC is shown in \autoref{fig:laser-eff}. The highest efficiency and highest middle point $\upmu$ is generally observed at the center of the pixels and decreases as the scan approaches the edges. This decrease is attributed to charge sharing, i.e., the division of electrons between two or more nearby pixels. The $\upsigma$ is uniform across the measured areas, giving a uniform noise characteristics. 

\begin{figure}
    \centering
    \begin{subfigure}{0.45\textwidth}
    \centering
    \includegraphics[trim=2.9cm 0 0 0,width=0.6\linewidth]{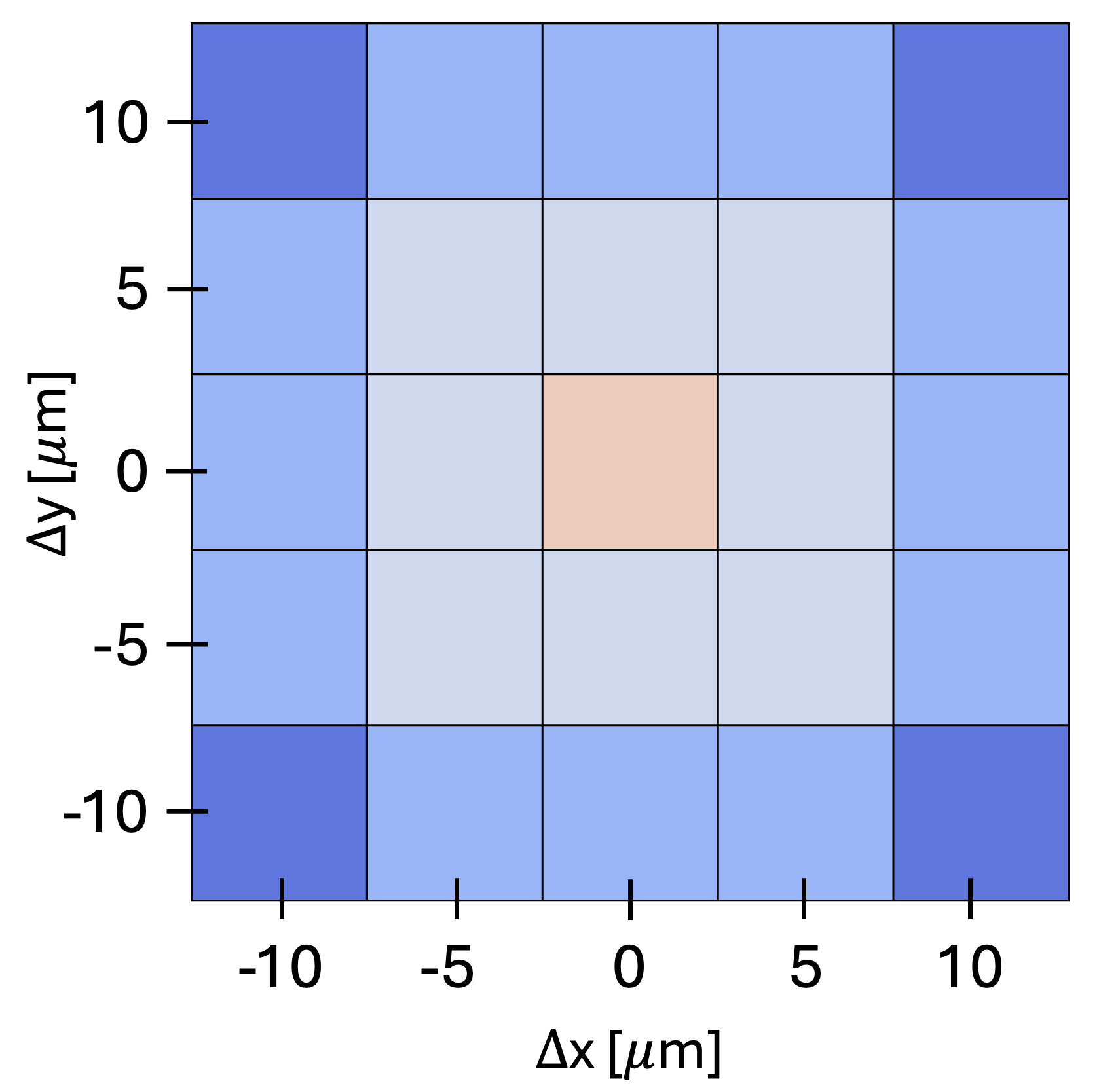}
    \caption{}
    \label{fig:laser.inpixel_scheme}
    \end{subfigure}
    \centering
    \begin{subfigure}{0.45\textwidth}
    \centering
    \includegraphics[width=\linewidth]{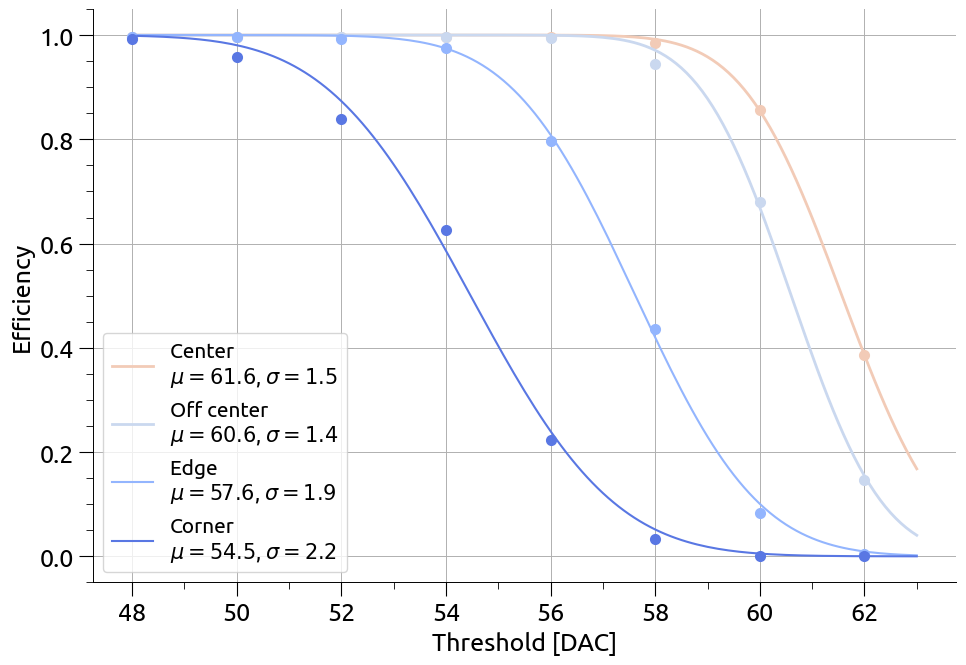}
    \caption{}
    \label{fig:laser.VCASN-fits}
    \end{subfigure}
    \caption{Scheme of the in-pixel scanned areas \subref{fig:laser.inpixel_scheme} and results of S-curve fits on scan points at four different in-pixel positions \subref{fig:laser.VCASN-fits}.}
\end{figure}

\begin{figure}[htbp]
    \centering
    \begin{subfigure}{0.45\textwidth}
    \centering
    \includegraphics[width=\linewidth]{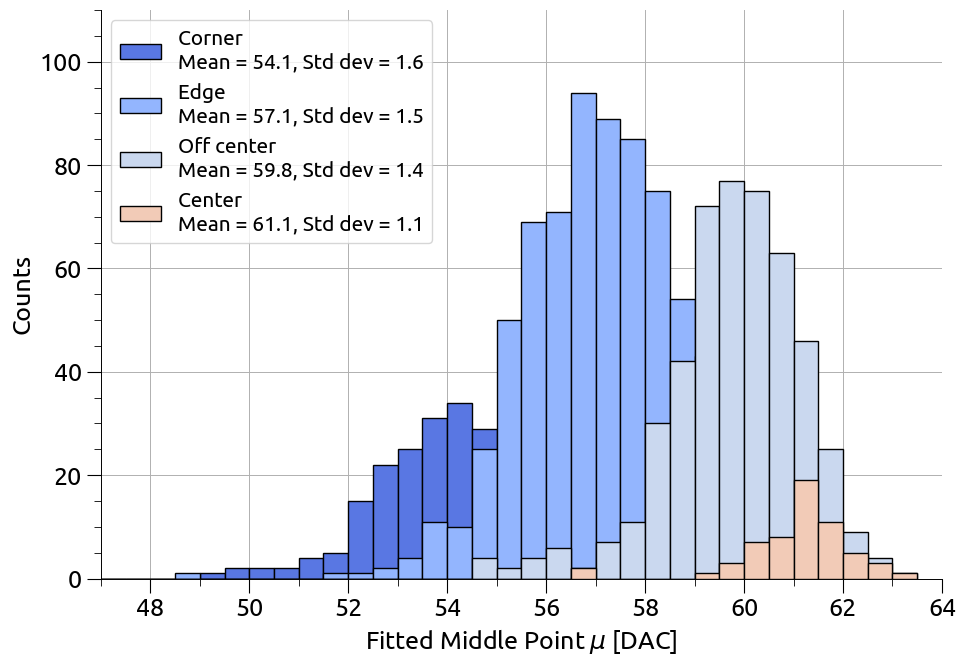}
    \caption{}
    \label{fig:laser.central-fits.mu}
    \end{subfigure}
    \begin{subfigure}{0.45\textwidth}
    \centering
    \includegraphics[width=\linewidth]{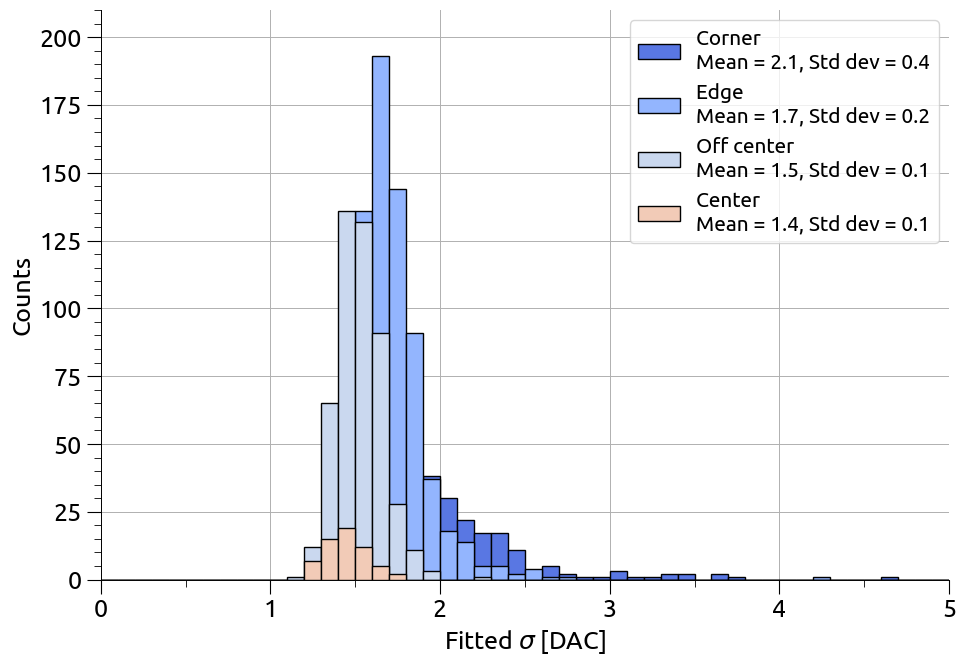}
    \caption{}
    \label{fig:laser.central-fits.sigma}
    \end{subfigure}
    \caption{ Distributions of the fitted S-curve parameters of $\upmu$ \subref{fig:laser.central-fits.mu} and $\upsigma$ \subref{fig:laser.central-fits.sigma} for $10 \times 6$ pixels in center of the ARCADIA chip. The color of the distributions distinguishes measurements from various positions where the laser scan is performed within a pixel.}
    \label{fig:laser.central-fits}
\end{figure}

The fitted regions are grouped into four categories: scans at the pixel center (center), scans adjacent to the center within $\pm 5$ \si{\micro\meter} (off-center), and scans at the pixel border (edge and corner), as depicted in the scheme in \autoref{fig:laser.inpixel_scheme}. \autoref{fig:laser.VCASN-fits} shows S-curve fits for the four different in-pixel positions. The resulting distributions of the fitted $\upmu$ and $\upsigma$ values are shown in \autoref{fig:laser.central-fits}. It can be observed that the middle point of the s-curve moves to lower values as the distance from the center of the pixel increases. The shift of the middle point comparing the pixel center with the corner corresponds to 7 DAC units. The relative standard deviation increases from around 1.8\% to 3\%. As regards the sigma of the S-curve, it increases as the distance from the pixel center increases, and the relative spread of the sigma distribution increases significantly from 7\% to 19\%. 
Overall, the pixels demonstrate uniform efficiency and response to the IR laser.

\section{Conclusions}
\label{sec:conclusions}

The characterization of the ARCADIA MD3 chip is presented in this work, demonstrating the functionality of the prototype. Chip threshold uniformity is evaluated using both test pulse and $^{55}$Fe measurements: the pattern observed in the threshold map with test pulsing does not appear in the $^{55}$Fe map, indicating that it is an artifact of the injection circuit. A conversion relation from DAC units to number of electrons in the high-threshold range, is established using both $^{55}$Fe and monochromatic fluorescence X-rays. At a threshold value of about 1700 electrons, the threshold dispersion is less than 8\% while considering all pixels on the chip. \\
Finally, precision scans with an IR-laser beam demonstrate that the sensor achieves full efficiency uniformly across the entire pixel matrix. As regards the in-pixel efficiency, the highest efficiency is observed at the pixel center; moving away from the center, the efficiency decreases. \\
The very good noise figure of about 30 electrons at room temperature ensures achieving a sufficient (factor 15 or better) signal-to-noise ratio already at depletion thicknesses of \SI{10}{\micro\meter}, i.e. with sensors thinned down to \SI{20}{\micro\meter}. Indeed, within the ARCADIA technology framework, the fabrication of very thin sensors is also feasible starting from a p+ substrate with the double epitaxial layer grown on top. \\ Together with the low power consumption and good threshold uniformity, this makes the ARCADIA technology an excellent stepping stone toward the realization of pixel detectors for the next generation High Energy Physics experiments, where sensors thinning and low power will be mandatory to meet the extremely low material budget requirements.
The tracking performance of the ARCADIA MD3 will be presented in forthcoming publications, reporting the results of a three-plane MD3 telescope tested with a \SI{120}{\giga\eV} proton beam at the FNAL test beam facility.


\acknowledgments
The author(s) declare that financial support was received for the research and/or publication of this article. This work has received funding from INFN CSN1 and CSN5 under the Call 2018 project ARCADIA. This study was carried out within the Space It Up and received funding from the ASI and the MUR – Contract n. 2024-5-E.0 - CUP n. I53D24000060005.
This work was produced by FermiForward Discovery Group, LLC under Contract No. 89243024CSC000002 with the U.S. Department of Energy, Office of Science, Office of High Energy Physics. Publisher acknowledges the U.S. Government license to provide public access under the DOE Public Access Plan \href{https://www.energy.gov/downloads/doe-public-access-plan}{DOE Public Access Plan}.
This work was supported by the Science Committee of Republic of Armenia (Research projects No.22AA-1C009 and 22rl-037).

\bibliographystyle{JHEP}
\bibliography{biblio}

\end{document}